\documentclass[a4paper]{article}

\usepackage[centertags]{amsmath}
\usepackage{amsfonts}
\usepackage{amssymb}
\usepackage{amsthm}

\usepackage{graphicx}
\usepackage{hyperref}
\usepackage{wrapfig}

\newcommand{\bea}{\begin{eqnarray}}
\newcommand{\eea}{\end{eqnarray}}

\newcommand{\be}{\begin{equation}\begin{gathered}}
\newcommand{\ee}{\end{gathered}\end{equation}}

\newcommand{\ben}{\begin{equation*}\begin{gathered}}
\newcommand{\een}{\end{gathered}\end{equation*}}

\makeatother%
\begin{document}

\title{Some symmetry group aspects of perfect plane plasticity system
}

\author{S.I. Senashov\thanks{sen@sibsau.ru},\\
              Siberian State Aerospace University\\
               31, Krasnoyarsky Rabochy Av., Krasnoyarsk, Russia, 660014
                \and
           A. Yakhno\thanks{alexander.yakhno@cucei.udg.mx},\\
             Departamento de Matem\'aticas, CUCEI, Universidad de Guadalajara,\\
            Blvd. Marcelino Garc\'ia Barrag\'an, 1421, C.P. 44430, Guadalajara, M\'exico\
 }

\date{}

\maketitle

\begin{abstract}
In this paper, all the known classical solutions of plane perfect plasticity system under  Saint Venant -- Tresca -- von Mises yield criterion are associated with some group of  point symmetries. The equations of slip-line families for all solutions are constructed, which permits to determine explicitly boundaries of plastic areas.

It is shown, how one can determine the compatible velocity solution for known stresses, considering symmetries. Some invariant solutions of velocities for Prandtl stresses are constructed. The mechanical sense of obtained velocity fields is discussed.
\end{abstract}


\begin{center}
\begin{quotation}
\centering
{\it To the blessed memory of our teacher D.D.~Ivlev}
\end{quotation}
\end{center}

\section*{Introduction}

Let us consider the well known system of two-dimensional perfect plasticity \cite{Kachanov:2004}:
\begin{equation}
\label{cartesian}\begin{gathered}
\frac{\partial\sigma}{\partial x} - 2k\left(\frac{\partial{\theta^c}}{\partial x}\cos{2\theta^c} + \frac{\partial\theta^c}{\partial y}\sin{2\theta^c}\right) = 0,\\
\frac{\partial\sigma}{\partial y} - 2k\left(\frac{\partial{\theta^c}}{\partial x}\sin{2\theta^c} - \frac{\partial\theta^c}{\partial y}\cos{2\theta^c}\right) = 0,
\end{gathered}
\end{equation}
which follows from equilibrium equations in the absence of body forces:
\be
\label{equil_c}
\frac{\partial \sigma_x}{\partial x} + \frac{\partial \tau_{xy}}{ \partial y} = 0, \ \frac{\partial \tau_{xy}}{\partial x} + \frac{\partial \sigma_y}{ \partial y} = 0
\ee
and yield Saint Venant -- Tresca -- von Mises criterion
\be
\label{yield_c}
(\sigma_x - \sigma_y)^2 + 4\tau_{xy}^2 = 4k^2
\ee
through the change of variables (due to L\'evi):
\be
\label{change}
\sigma_x = \sigma - k \sin2\theta^c,
\sigma_y = \sigma + k \sin2\theta^c,
\tau_{xy} = k \cos 2\theta^c.
\ee
 Here $\sigma_x$, $\sigma_y$ are the normal components and $\tau_{xy}$ the tangential component of stress in a rectangular coordinates $(x,y)$, $\theta^c + \pi/4$ is a slope angle of the maximum principal stress relative to the axis $Ox$, $\sigma$ is the mean compressive stress (or hydrostatic pressure), and $k$ is a constant of the material.  
 
To construct the velocity field, which is consistent with solution of (\ref{cartesian}) of the form $\sigma=\sigma_0(x,y)$, $\theta^c = \theta^c_0(x,y)$ one needs to solve the following system of linear equations:
\be
\label{uv_cartesian}
\left( \frac{\partial u}{\partial y} + \frac{\partial v}{\partial x} \right) \sin 2\theta_0 + \left( \frac{\partial u}{\partial x} - \frac{\partial v}{\partial y} \right) \cos 2\theta_0 = 0, \\
\frac{\partial u}{\partial x} + \frac{\partial v}{\partial y} = 0,
\ee
where $u(x,y)$, $v(x,y)$ are the components of velocity vector.

 System (\ref{cartesian}) describes the stress state of material, which is being plastically deformed and has been studied over one hundred years. The main contribution in determining closed-form solutions was made by L.~Prandtl and A.~Nadai at the beginning of the 20th century. These solutions are now known as classical ones and serve as a good approximations for the real mechanical processes. 
 
The symmetry method of solving differential equations was applied to system (\ref{cartesian}) for the first time in \cite{Annin:1978}. In \cite{Annin:1985} the main part of symmetries for system (\ref{equil_c}), (\ref{yield_c}) was calculated and some known solutions were associated with correspondent transformation groups. Finally, the complete Lie algebra of all admissible point symmetries of (\ref{cartesian}) was determined in \cite{Senashov:1988} and all conservation laws as well Lie-Backlund symmetries were constructed. Moreover, in the series of papers \cite{Senashov:2007}, \cite{NA:2009}, \cite{Yakhno:2010}  point transformation groups were used to deform some known solutions and in \cite{Yakhno:2012} conservation laws were applied to solve the main boundary problems. 

Let us note, that a group analysis of the differential equations, is a semi-inverse method. It means that firstly some solution should be found, then one can determine the corresponding boundary conditions. It allows to solve statically determined problems for (\ref{cartesian}), (\ref{uv_cartesian}), i.e. when boundary conditions involve only the stresses, sufficient to permit a determination of the plastic region and state of stress without considering the velocities. The velocities can be calculated afterwards.

The interest to the system of plane plasticity has been recently renewed. In \cite{Lamothe:JMATH} the known symmetries, admitted by (\ref{cartesian}), (\ref{uv_cartesian}) were used to determine some solutions in the form of a propagation wave. In \cite{Lamothe:2012} the complete Lie algebra of symmetries for (\ref{cartesian}), (\ref{uv_cartesian}) is calculated. Two new operators of symmetries are determined. 

In work \cite{Hlavach_Marvan:2013} the plastic stress states of the round sphere are analyzed from the point of view of orthogonal equiareal patterns, and an interesting relation between slip-line fields on the sphere and sine-Gordon equation is shown.
 
The main goal of the present work is, on the one hand, to re-consider the relation between the known classical solutions and the group of point transformations (symmetries), because some of results were omitted previously or were not published. On the other hand, we analyze velocities, which are compatible with stresses, defined by (\ref{cartesian}) from the symmetry point of view.

The algorithm of construction of invariant solutions is well known. An interested reader can find a numerous bibliography on this matter (see, for example, \cite{Ovsiannikov:1982}). The main steps are: 1) for the given system define a Lie algebra of admissible operators, generating a group of point continuos transformations (symmetries); 2) determine the optimal system on non-similar subalgebras; 3) construct a set of invariants for representatives of subalgebras and if possible, find out the form of invariant solution; 4) substituting the invariant solution form into the original system, the so-called factor-system is obtained, which is in our case the system of ordinary differential equations; 5) solving a factor-system one can obtain the invariant solution.
 
The paper is organized as follows. In Section \ref{sec:notes} we consider some known solutions, obtained through heuristic methods and their mechanical interpretation. Section \ref{sec:group} is devoted to correspondence between symmetries and known classic closed-form solutions. Finally, the construction of velocity field, related to stress solution is discussed in Section \ref{sec:velocity}. 

The following information will be useful. System (\ref{cartesian}) is a hyperbolic one. Two families of characteristic lines are defined by two equations:
\be
\label{char_cc}
\frac{dy}{dx} = \tan \theta^c, \  \frac{dy}{dx} = -\cot \theta^c,
\ee
and corresponding Riemann invariants, which are constant along characteristics, look like
\be
\label{Riemmann}
\frac{\sigma}{2k} - \theta^c = \xi, \ \frac{\sigma}{2k} + \theta^c = \eta.
\ee
In the theory of plane perfect plasticity, characteristic lines coincide with so-called slip lines: curves whose directions at every point coincide with those of the maximum shear strain-rate. Equations (\ref{char_cc}) define slip-line field. As indicated in \cite{Hill:1950}, the field of slip-lines is the fundamental unknown element to be determined. 

In polar coordinates $(r, \varphi)$, system (\ref{equil_c}), (\ref{yield_c}) has the form
\ben
\frac{\partial \sigma_r}{ \partial r} + \frac{1}{r} \frac{\partial \tau_{r\varphi} }{ \partial \varphi} + \frac{\sigma_r - \sigma_\varphi}{r} = 0, \\
 \frac{\partial \tau_{r\varphi} }{ \partial r} + \frac{1}{r} \frac{\partial \sigma_\varphi}{ \partial \varphi} + \frac{2}{r} \tau_{r\varphi} = 0, \\
(\sigma_r - \sigma_\varphi)^2 + 4\tau_{r\varphi}^2 = 4k^2,
\een
where $\sigma_r$, $\sigma_\varphi$ are the radial and angular components and $\tau_{r\varphi}$ the tangential component of stress.

Introducing in a similar way
\ben
\sigma_r = \sigma - k\sin 2\theta^p,\
\sigma_\varphi = \sigma + k\sin 2\theta^p,\ 
\tau_{r\varphi} = k \cos 2\theta^p 
\een
 into above system, one can obtain:
\be
\label{polar}
r \frac{\partial \sigma^p}{\partial r} - 2k \left( r \frac{\partial \theta^p}{\partial r} \cos 2\theta^p +  \frac{\partial \theta^p}{\partial \varphi} \sin 2\theta^p\right) = 2k \sin 2\theta^p \\
\frac{\partial \sigma^p}{\partial \varphi} - 2k \left( r \frac{\partial \theta^p}{\partial r} \sin 2\theta^p -  \frac{\partial \theta^p}{\partial \varphi} \cos 2\theta^p\right) = - 2k \cos 2\theta^p,
\ee
where $\sigma^p = (\sigma_r + \sigma_\varphi)/2 = \sigma$ and $\theta^p$ is an angle between radio and slope of the slip-line.

Two families of characteristic lines are defined by following equations:
\be
\label{char_polar}
\frac{dr}{d\varphi} = - r \tan \theta^p, \  \frac{dr}{d\varphi} =  r \cot \theta^p.
\ee

From system (\ref{polar}), using (\ref{change}) and $\theta = \theta^c = \theta^p + \varphi$, we have:
\be
\label{polar2}
r \frac{\partial \sigma}{\partial r} - 2k \left( r \frac{\partial \theta^c}{\partial r} \cos 2(\theta^c-\varphi) +  \frac{\partial \theta^c}{\partial \varphi} \sin 2(\theta^c-\varphi) \right) = 0  \\
\frac{\partial \sigma}{\partial \varphi} - 2k \left( r \frac{\partial \theta^c}{\partial r} \sin 2(\theta^c-\varphi) -  \frac{\partial \theta^c}{\partial \varphi} \cos 2(\theta^c-\varphi) \right) = 0.
\ee

\section{Some notes about known solutions}\label{sec:notes}


\subsection{Revuzhenko solution}

Revuzhenko in his paper \cite{Revuzhenko:1975} considered the so called limiting equilibrium equations:
\bea
\label{sigma_phi}
\frac{1}{2k}\frac{\partial \sigma}{\partial \lambda_1} - \frac{\partial \phi}{\partial \lambda_1} =0, \  \frac{1}{2k} \frac{\partial \sigma}{\partial \lambda_2} + \frac{\partial \phi}{\partial \lambda_2} =0,
\eea
\bea
\label{xy_phi}
\frac{\partial y}{\partial \lambda_1} =\tan\left(\phi - \frac{\pi}{4} \right) \frac{\partial x}{\partial \lambda_1} , \  \frac{\partial y}{\partial \lambda_2} =\tan\left(\phi + \frac{\pi}{4} \right) \frac{\partial x}{\partial \lambda_2},
\eea
where $\phi = \theta^c + \pi/4$ and $\lambda_1 = \eta$, $\lambda_2 = \xi$ are characteristic coordinates (\ref{Riemmann}). Let us take $2k=1$, then the first two equations (\ref{sigma_phi}) give
\be
\label{FG_sigma_theta}
\sigma - \theta^c = F(\xi), \ \sigma + \theta^c = G(\eta),
\ee
and equations (\ref{xy_phi}) obtain the following form
\bea
\label{char_c}
\frac{\partial y}{\partial \eta} =\tan\theta^c \frac{\partial x}{\partial \eta} , \  \frac{\partial y}{\partial \xi} =-\cot \theta^c \frac{\partial x}{\partial \xi},
\eea
where $\theta^c = (G-F)/2$, $G(\eta)$ and $F(\xi)$ are arbitrary functions and  not identical constants.

Solving linear equations (\ref{char_c}), closely related to (\ref{char_cc}), one can find parametric equations $x = x(\xi,\eta)$, $y = y(\xi,\eta)$ of characteristic lines in Cartesian coordinates. If the slip-line field is known, then one can calculate functions $\sigma(\xi,\eta)$, $\theta(\xi,\eta)$ at the point $(x,y)$ using (\ref{FG_sigma_theta}).

In polar coordinates, system (\ref{char_c}) takes the form
\bea
\label{char_p}
\frac{1}{r} \frac{\partial r}{\partial \eta} = \cot(\theta^c - \varphi) \frac{\partial \varphi}{\partial \eta},\ 
\frac{1}{r} \frac{\partial r}{\partial \xi} = -\tan(\theta^c - \varphi) \frac{\partial \varphi}{\partial \xi},
\eea
and is similar to (\ref{char_polar}).

 Eliminating $r$ from the above system, we obtain one equation for the function $u = \tan\theta^p$:
\be
\label{eq_u}
\frac{\partial^2}{\partial \xi \partial \eta} \ln |u| - \frac{G^\prime}{2} \frac{\partial }{\partial \xi} \frac{1}{u} + \frac{F^\prime}{2} \frac{\partial }{\partial \eta} u  = 0.
\ee
Its solution can be determined by separating variables and has the following form ($a,b,c = const.$)
\be
\label{u_solution}
u^2 = \tan^2 \theta^p = \frac{a G(\eta) + b}{a F(\xi) + c}.
\ee
Then, for the simplicity, one can take
\be
\label{GF}
\tan^2 \theta^p = \frac{\eta^2}{\xi^2}, \ G = 2(\eta^2 - \pi/8),\ F = 2(\xi^2 + \pi/8).
\ee
Regressing to variable $\varphi = \theta^c - \theta^p = (G - F)/2 - \theta^p$ we have:
\be
\label{sol_phi}
\varphi = \eta^2 - \xi^2 - \pi/4 \mp \arctan \frac{\eta}{\xi}. 
\ee
The sign $\mp$ follows from the quadratic type of (\ref{u_solution}). 

Taking $\tan\theta^p = \pm \eta/\xi$ and integrating (\ref{char_p}), 
which take the form:
\bea
\label{char_p_xi_eta}
\frac{1}{r} \frac{\partial r}{\partial \eta} = \pm \frac{\xi}{\eta} \frac{\partial \varphi}{\partial \eta},\ 
\frac{1}{r} \frac{\partial r}{\partial \xi} = \mp \frac{\eta}{\xi} \frac{\partial \varphi}{\partial \xi}, 
\eea
one can determine function $r$
\be
\label{sol_r_p}
r = e^{\pm 2\xi\eta} \sqrt{\xi^{-2} + \eta^{-2}}.
\ee
Solution for $\sigma$, $\theta$ is given by functions $G$, $F$:
\be
\label{sol_sigma}
\sigma = \frac{G + F}{2} = \xi^2 + \eta^2, \ \theta^c = \frac{G - F}{2} = \eta^2 - \xi^2 - \frac{\pi}{4}.
\ee

From the necessary condition for an envelope of a family of curves $\dfrac{\partial (r,\varphi)}{\partial (\xi, \eta)} = 0$, taking into account equations (\ref{char_p_xi_eta}), we have
$
\dfrac{\partial \varphi}{\partial \xi} \dfrac{\partial \varphi}{\partial \eta} = 0,
$
or
\ben
2 \xi \mp \frac{\eta}{\xi^2 + \eta^2} = 0, \ 
2 \eta \mp \frac{\xi}{\xi^2 + \eta^2}= 0. 
\een

Let us note, that two solutions can be obtained one from the other through the change $\eta \leftrightarrow - \eta$ (or $\xi \leftrightarrow -\xi$), the same is valid for conditions for envelope lines. So there is no principal difference between mechanical interpretations of these two solutions.

Let us consider solution (\ref{sol_phi}), (\ref{sol_r_p}) with the lower sign.
Adding the condition 
\be
\label{phi_xi}
-\frac{\partial \varphi}{\partial \xi} = 2 \xi + \frac{\eta}{\xi^2 + \eta^2} = 0
\ee
and expressing $\eta = \eta(\xi)$, one can obtain the corresponding envelope for the family of characteristic curves $\xi = const.$ Corresponding envelope for the second family of characteristics is defined by condition
\be
\label{phi_eta}
\frac{\partial \varphi}{\partial \eta} = 2 \eta + \frac{\xi}{\xi^2 + \eta^2} = 0.
\ee
Let us note, that the above equation through the change $\xi \leftrightarrow \eta$ is reduced to (\ref{phi_xi}) and it defines cusp (return points) curve for the family $\xi = const.$ Moreover, from relations (\ref{phi_xi}) and (\ref{phi_eta}) it follows that along an envelope, the product $\xi\eta$ should be negative.

\begin{figure}[h]
\begin{center}
\includegraphics[width=0.5\linewidth]{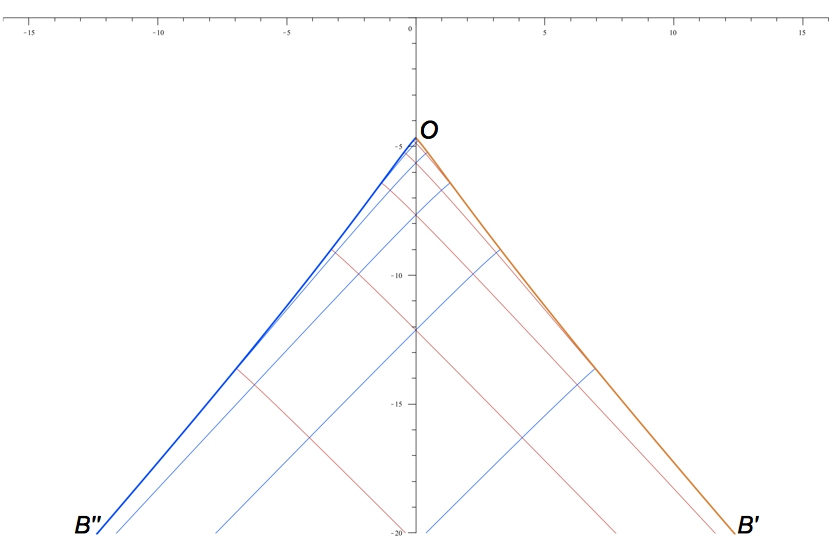}
\caption{Angle area of the first kind}
 \label{fig:1}
\end{center}
\end{figure}

\paragraph{1}
Let us consider case (\ref{phi_xi}),  when $\eta \in (-1/2, 0)$, $\xi \in (0, 1/2)$. This case corresponds to envelope $OB'$ (see Fig. \ref{fig:1}). In a similar way one can obtain envelope $OB''$, taking the root of (\ref{phi_eta}) and considering $\eta \in (0,1/2)$, $\xi \in (-1/2, 0)$.

Right angle plastic area, bounded by curves $OB'$ and $OB''$ is, in some sense, an analogy of Nadai solution for the converging channel with straight line borders. Point $O$ is the singular one.

\begin{figure}[h]
\begin{center}
\begin{minipage}[h]{0.49\linewidth}
\includegraphics[width=1\linewidth]{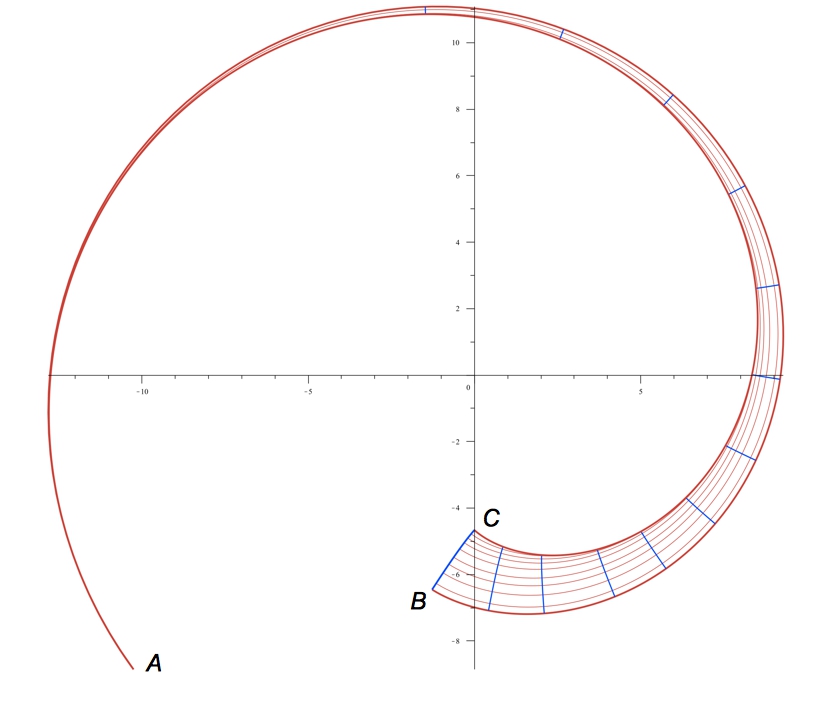}
\caption{Hornlike area}
\label{fig:2}
\end{minipage}
\hfill
\begin{minipage}[c]{0.45\linewidth}
 \includegraphics[width=1\linewidth]{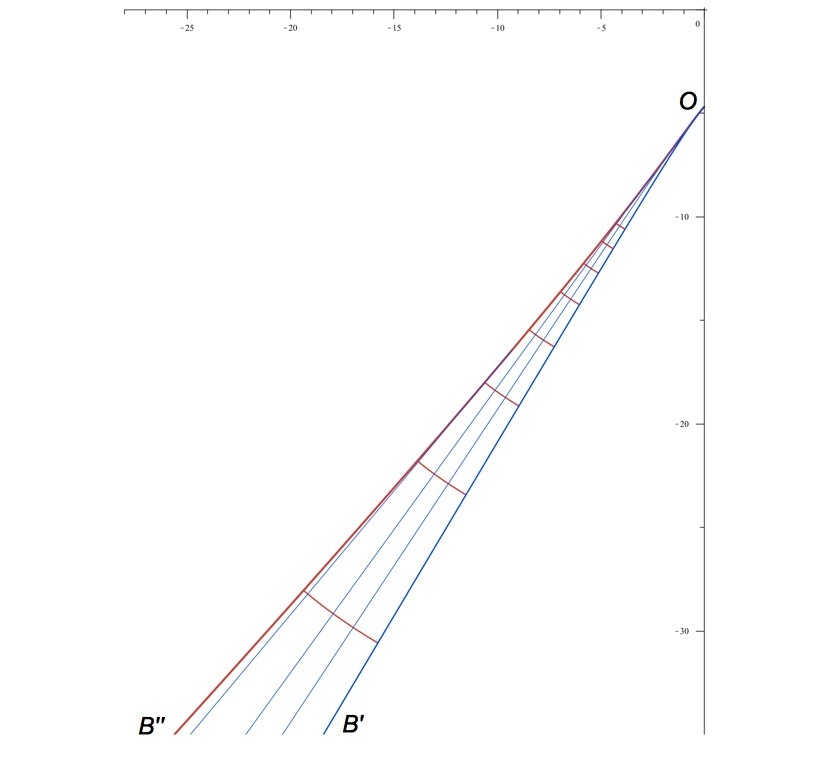}
 \caption{Angle area of the second kind} 
 \label{fig:3}  
 \end{minipage}
\end{center}
\end{figure}

\paragraph{2}
Taking the root of (\ref{phi_xi}), namely $\eta(\xi) = (-1 - \sqrt{1-16\xi^4})/(4\xi)$, one obtain the envelope $AC$ (see Fig. \ref{fig:2}), which is a part of a spiral. Another boundary $AB$ of this hornlike plastic area is one of the slip lines $-1/2 <\xi = \xi_0 < 0$.  The contact of curves is achieved at point $A$. 

\paragraph{3}
Let us consider envelope  $OB''$ of the family $\eta = const.$ (Fig. \ref{fig:3}), obtained from (\ref{phi_eta}), taking $\xi(\eta) = (-1 + \sqrt{1-16\eta^4})/(4\eta)$ and $\eta \in (0, 1/2)$, $\xi \in (-\infty, -1/2)$. Another boundary $OB'$ of the angle area is the slip line $\eta = 1/2$. 

For more details for boundary conditions see \cite{Revuzhenko:1975}.

\subsection{Nadai solution for two concentric circles}

In \cite{Nadai:1928} Nadai proposed the form of dependence $\tau_{r\varphi} = h(r) = k \cos 2\theta^p$, that  from (\ref{polar}) gives:
\be
\label{Nadai_two_circ}
\cos 2\theta^p = C_1 r^{-2} + C_2, \ \sigma = -2kC_2 \varphi + f(r),
\ee
where $f(r)$ can be determined by quadrature from equation
\be
\label{ff}
f' -2k (\theta^p)' \cos 2\theta^p = 2k r^{-1} \sin 2\theta^p.
\ee
This solution is interpreted as stresses in the area, bounded by two concentric circles $r=a$, $r=b$, so that
\ben
 \tau_{r\varphi}|_{r=a} = -k,\  \tau_{r\varphi}|_{r=b} = k,
\een
which gives
\be
\label{cc}
C_1 = - 2 \frac{a^2 b^2}{b^2 - a^2}, \ C_2 =  \frac{a^2 + b^2}{b^2 - a^2}.
\ee

For simplicity, let us take $a=1$, $b=\sqrt{2}$, then, taking into account that
\ben
\tan \theta^p = \frac{1}{\sqrt{2}} \sqrt{\frac{2-r^2}{r^2-1}}
\een
from the first equation for slip-lines (\ref{char_polar}) one can obtain the relation for the first family ($1\leqslant r \leqslant \sqrt{2} $)
\ben
 \frac{1}{\sqrt{2}} \arctan  \frac{3 r^2 - 4}{2\sqrt{2} \sqrt{(r^2 - 1)(2-r^2)}} - \arcsin(2 r^2 - 3) = \sqrt{2} \varphi + K_1,
\een
which are parts of epicycloids.
 
The second equation gives the equation of the second family of slip-lines (hypocycloids):
\ben
{\sqrt{2}} \arctan \frac{3 r^2 - 4}{2\sqrt{2} \sqrt{(r^2 - 1)(2-r^2)}}  - \arcsin(2 r^2 - 3) = 2 \sqrt{2} \varphi + K_2.
\een

\begin{figure}[h]
\begin{center}
\includegraphics[width=0.45\linewidth]{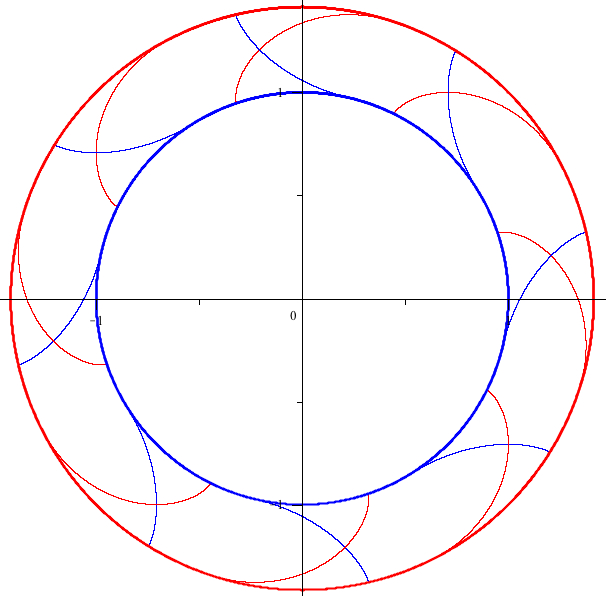}
\caption{Epi- and hypocycloids}
 \label{fig:hypo}
\end{center}
\end{figure}

Let us note, that concentric circles are envelopes for corresponding families of epi- and hypocycloids (see Fig. \ref{fig:hypo}).

\section{Group analysis and invariant solutions for stresses}\label{sec:group}

It is known \cite{Senashov:1988} that system (\ref{cartesian}) admits an infinite algebra of generalized (highest) symmetries. The basis of Lie algebra $L_{\sigma\theta}$ of point transformations is formed with the following operators:
\begin{equation}
\begin{gathered}
X_1 = x \frac{\partial}{\partial x} + y\frac{\partial}{\partial y}, \ X_2 = -y \frac{\partial}{\partial x} + x \frac{\partial}{\partial y} + \frac{\partial}{\partial \theta}, \ X_3 = \frac{\partial}{\partial \sigma},\\
 X_4 = \xi_1(x,y,\sigma,\theta) \frac{\partial}{\partial x} + \xi_2(x,y,\sigma,\theta) \frac{\partial}{\partial y} - 4 k\theta \frac{\partial}{\partial \sigma}-\frac{\sigma}{k}\frac{\partial}{\partial \theta}, \\
X_5 = \xi(\sigma,\theta)\frac{\partial}{\partial x} + \eta(\sigma,\theta) \frac{\partial}{\partial y},
\label{eq3}
\end{gathered}
\end{equation}
where
\[
\xi_1 = x\cos 2\theta + y\sin 2\theta + y\frac{\sigma}{k},\ 
\xi_2 = x\sin 2\theta - y\cos 2\theta - x\frac{\sigma}{k},
\]
and $(\xi,\eta)$ is an arbitrary solution of the linear system
\begin{equation}
\begin{gathered}
\frac{\partial x}{\partial \theta} - 2k\left(\frac{\partial x}{\partial \sigma} \cos 2\theta + \frac{\partial y}{\partial
\sigma} \sin 2\theta\right) = 0, \\
\frac{\partial y}{\partial \theta} - 2k\left(\frac{\partial x}{\partial\sigma} \sin 2\theta - \frac{\partial y}{\partial\sigma} \cos 2\theta\right) = 0,
\label{linear}
\end{gathered}	
\end{equation} 
obtained from (\ref{cartesian}) applying hodograph transformations of the form $x = x(\sigma,\theta)$, $y = y(\sigma,\theta)$ and assuming that Jacobian of transformation is not equal to zero. Operator $X_5$ forms infinite-dimensional subalgebra of $L_{\sigma\theta}$ and we consider it later (see subsection \ref{sub:infinite}). 

In polar coordinates above operators (except $X_5$) take the following form:
\begin{equation}
\begin{gathered}
X_1^p = r \frac{\partial}{\partial r}, \ X_2^p = \frac{\partial}{\partial \varphi} +  \frac{\partial}{\partial \theta}, \ X_3^p = \frac{\partial}{\partial \sigma},\\
 X_4^p = r\cos 2(\theta - \varphi) \frac{\partial}{\partial r} + \left(\sin 2(\theta - \varphi) - \frac{\sigma} k \right) \frac{\partial}{\partial \varphi} - 4 k\theta \frac{\partial}{\partial \sigma}-\frac{\sigma}{k}\frac{\partial}{\partial \theta}.
\label{oper_polar}
\end{gathered}
\end{equation}

Non-zero commutators of (\ref{eq3}) are as follows
\be
\label{comm_sigma}
[X_2, X_4] = - 4k X_3, \ [X_3, X_4] = -X_2/k.
\ee

From the symmetry point of view, it is necessary to construct the so called optimal system of non similar one-dimensional subalgebras \cite{Ovsiannikov:1982} in order to define different exact solutions. The optimal system of one-dimensional subalgebras for finite part of $L_{\sigma\theta}$ is the following one ($\alpha \in \mathbb{R}$):
\be
\Theta_1 = \left< X_4 + \alpha X_1 \right> = \left< X_4 + \alpha r\partial_r \right>,  \
\Theta_2 = \left< X_3 + \alpha X_1 \right> = \left< \partial_\sigma + \alpha r\partial_r  \right>, \\
 \Theta_3 = \left< X_2 + \alpha X_1 \right> = \left< \partial_\varphi + \partial_\theta + \alpha r\partial_r \right>,  \\ 
 \Theta_4^{(\pm)} = \left< X_3 \pm X_2 + \alpha X_1 \right> = \left< \partial_\sigma \pm \left(\partial_\varphi + \partial_\theta \right) + \alpha r\partial_r \right>, \
 \Theta_5 = \left< X_1 \right> =  \left<  r\partial_r \right>,
\ee
here we use $\partial_t = \dfrac{\partial}{\partial t}$ for the simplicity. Non similar subalgebras correspond to different values of $\alpha$. 

Let us note, that system (\ref{cartesian}) is invariant with respect to discrete symmetries:
\ben
x\to - x,\ \sigma \to - \sigma,\ \theta \to - \theta;\\
y\to - y,\ \sigma \to - \sigma,\ \theta \to - \theta,
\een
and systems (\ref{polar}) and (\ref{polar2}) admit transformations
\bea
&\varphi \to - \varphi,\ \sigma \to - \sigma,\ \theta^{p,c} \to - \theta^{p,c}; \label{reflection3} \\
&r\to - r \nonumber
\eea
respectively.

The transformation group, corresponding to operator $X_4$ has been calculated and called ''quasi-scale'' transformation in \cite{NA:2009}. 

\subsection{$\Theta_1$: quasi-scale transformation}

From the group analysis point of view, equation (\ref{eq_u}) admits the following symmetries:
\ben
Y_1 = - \frac{F}{F'} \partial_{\xi} + \frac{G}{G'} \partial_{\eta} + u\partial_u,\  Y_2 = \frac{2}{F'}\partial_{\xi}, \ Y_3 = \frac{2}{G'} \partial_{\eta},
\een
and there is no extension of the group for any specific form of $G$ and $F$. For  functions $G$, $F$, taken as in (\ref{GF}), above operators look like
\ben
Y_1 = - \xi \partial_{\xi} + \eta \partial_{\eta} + 2 u\partial_u,\  Y_2 = {\xi^{-1}}\partial_{\xi}, \ Y_3 = {\eta^{-1}} \partial_{\eta}.
\een

Invariant solution corresponding to operator $Y_1$ has the form $u = \eta^2 f(z)$, where $z = \xi \eta$. Substituting this form into (\ref{eq_u}) one can obtain its general solution in terms of Bessel functions. In particular, $f = \pm 1/z$ is one of the solutions, therefore solution (\ref{GF}) $u = \pm \eta/\xi$ is the particular one and is invariant with respect to $Y_1$.

Taking into account solutions (\ref{sol_phi}), (\ref{sol_r_p}), (\ref{sol_sigma}), we can express operator $Y_1$ in terms of $r$, $\varphi$, $\theta^c$, $\sigma$. Namely, for the higher sing solution $\tan\theta^p = \eta/\xi$ we have:
\bea
&&2u\partial_u = -2\xi \partial_\xi + 2\eta \partial_\eta, \nonumber \\
&&\partial_\xi = r \left( 2\eta - \frac{\eta^2}{\xi\left(\xi^2 + \eta^2\right)} \right) \partial_r + \left(-2\xi + \frac{ \eta}{\xi^2 + \eta^2} \right) \partial_\varphi + 2\xi\partial_\sigma - 2\xi\partial_{\theta^c}, \nonumber \\
&&\partial_\eta= r \left( 2\xi - \frac{\xi^2}{\eta\left(\xi^2 + \eta^2\right)} \right) \partial_r + \left( 2\eta - \frac{ \xi}{\xi^2 + \eta^2} \right) \partial_\varphi + 2\eta\partial_\sigma + 2\eta\partial_{\theta^c}. \nonumber
\eea
Taking into account, that 
\ben
\cos 2\theta^p = \frac{1-\tan^2 \theta^p}{1+\tan^2 \theta^p} = \frac{\xi^2 - \eta^2}{\xi^2 + \eta^2}, \nonumber \
\sin 2\theta^p = \frac{2\tan \theta^p}{1 + \tan^2 \theta^p} = \frac{2\xi\eta}{\xi^2 + \eta^2},
\een
we finally obtain
\ben
\frac{1}{3} Y_1  = - r\cos2\theta^p \partial_r + \left[ 2\sigma - \sin 2\theta^p \right]\partial_\varphi + 2\theta^c\partial_\sigma + 2\sigma\partial_{\theta^c} + \frac{\pi}{2} \partial_\sigma = - X_4^p + \frac{\pi}{2} X_3^p.
\een 
For the solution with lower sign, the operator $Y_1$ is transformed by analogy. 

That is why one can associate Revuzhenko solution with sub algebra $\Theta_1$, when $\alpha = 0$. 
The case when $\alpha \ne 0$ will be considered in the future work.

\subsection{$\Theta_2$: $\sigma$ - translation and scale}

Let us take $\alpha = -1/(2kc) \ne 0$, then invariants of $\Theta_2$ are: $I_1 = \varphi$, $I_2 = \theta$, $I_3 = \sigma - 1/\alpha \ln r$ and solution has the form
\be
\label{nadai_inv_t2}
\sigma = -2 k c\, \ln r + f(\varphi),\ \theta = \theta(\varphi).
\ee
Substituting the above form into (\ref{polar2}), one can obtain the following system of ordinary differential equations:
\bea
\label{nadai_factor_system}
\theta' \sin 2(\theta - \varphi) = -c, \ f' = 2kc \cot 2(\theta - \varphi).
\eea 
Integrating the first equation one can find the expression for $\theta - \varphi$, then integrating the second equation, we obtain ($c_i \in \mathbb{R}$)
 \bea
\label{Nadai_channel}
 && \sigma = -2kc \ln r - kc \ln\left[c + \sin 2(\theta - \varphi)\right] +  const,\\
&&\theta = \varphi - \arctan\left(\frac{\sqrt{c^2 - 1}}{c}\tan\left[\frac{\sqrt{c^2 - 1}}{c}\left(\theta + c_1\right)\right]- \frac{1}{c}\right),\ c^2 >1,\label{Nadai_channel_cm1}\\
&&\theta = \varphi + \arctan\left(\frac{\sqrt{1-c^2}}{c}\tanh\left[\frac{\sqrt{1-c^2}}{c}\left(\theta + c_2\right)\right] + \frac{1}{c}\right),\ c^2 < 1. \label{Nadai_channel_cl1}
\eea

Slip lines equations are the following ones:
\ben
r = k_1 \frac{e^{-\theta/c}}{\sqrt{c + \sin 2(\theta - \varphi)}}, \ r = k_2 \frac{e^{\theta/c}}{\sqrt{c + \sin 2(\theta - \varphi)}}, \ k_i\in \mathbb{R},
\een
and $\varphi = \varphi(\theta)$ is defined by (\ref{Nadai_channel_cm1}) or (\ref{Nadai_channel_cl1}).

The value of the constant $c$ is important for mechanical interpretation of stresses.

In the case $c>1$, the families of slip lines have the envelopes, and they represent the well-known Nadai solution \cite{Hill:1950}, \cite{Nadai:1924}  and describes the flow of plastic material through the wedge-shaped converging channel (total angle $2\alpha$)
\ben
\alpha + \frac{\pi}{4} = \frac{c}{\sqrt{c^2 - 1}}\arctan\sqrt{\frac{c + 1}{c - 1}},\ \alpha\in\left(0,\ \frac{\pi}{2}\right).
\een
Two straight lines $\varphi = \pm \alpha$ are boundaries of the channel. For more details about this solution see \cite{Yakhno:2010}.

In the case $c^2<1$, there is not any envelope for  slip line families. But solution (\ref{Nadai_channel}), (\ref{Nadai_channel_cl1}) can still be used to describe a plastic state of the channel with the shear stress $\tau_{r\varphi} = k \cos 2\left(\theta^{(1,2)}_0 - \varphi^{(1,2)}_0\right)$ along straight-line borders $\varphi = \varphi^{(1,2)}_0 = const.$, because $\theta^p = const.$ along the straight line $\varphi = const.$  The relation between $\theta^{(1,2)}_0$ and $\varphi^{(1,2)}_0$ is given by (\ref{Nadai_channel_cl1}).

Unfortunately, a new partially invariant solution for the angle $\theta$ announced in \cite{Lamothe:2012} coincides with the above well-known Nadai solution for the channel. Compatible velocity solution can be found, for example, in \cite{Hill:1950}.

In the case of $c=1$, from (\ref{nadai_factor_system}) and after applying (\ref{reflection3}), we have a singular solution of the form
\be
\label{Nadai_circ}
\theta = \varphi + \frac{\pi}{4}, \ \sigma = 2k \ln \frac{r}{R} + k - p,
\ee
which is a well known Nadai solution describing plastic state around a circular cavity of radius $R$, situated in an infinite medium loaded by uniformly distributed pressure $p$, with the tangential stress equal to zero. For more details about this solution see \cite{NA:2009}. Let us note, that solution, obtained in \cite{Lamothe:2012} as  partially invariant solution, corresponding to the subalgebra generated by operator $K\sim X_4$, coincides with (\ref{Nadai_circ}). As for the velocity solution of (\ref{uv_cartesian}), compatible with Nadai solution see, for example, \cite{Sokolovsky:1950}.

The non singular solution of (\ref{nadai_factor_system}) is
\ben
\sigma = -2k \ln r - k \ln \left[1 + \sin 2(\theta - \varphi) \right] + const., \ \varphi = \theta +  \arctan \left( 1 + \frac{1}{\theta - A}\right), \ A \in \mathbb{R}.
\een

The case $\alpha = 0$ does not produce any invariant solution, because the necessary condition of existence of invariant solution \cite{Ovsiannikov:1982} is not satisfied.

\subsection{$\Theta_3$: rotation and scale}

a) The case $\alpha = 0$ corresponds to the rotation subgroup $\left< \partial_\varphi + \partial_{\theta} \right>$ and has invariants: $I_1 = r$, $I_2 = \sigma$, $I_3 = \theta - \varphi$, then invariant solution is as follows
\ben
\sigma = f(r), \ \theta = \varphi + g(r).
\een
After substituting into (\ref{polar2}) we have:
\be
\label{nadai_3}
r f' = \frac{2k}{\sin 2g}, \ rg' = \cot 2g.
\ee
The second equation gives two solutions
\ben
\cot g = \pm \frac{\sqrt{C^2 r^4 - 1}}{Cr^2 - 1}.
\een
Taking the boundary condition at $r=R$ as $\theta(R) = \varphi - \pi/2$ we obtain $C^2 = 1/R^4$ and $\cos 2g = -R^2/r^2$, so $C = - 1/R^2$. Then $\sin 2g = \pm \sqrt{r^4 - R^4}/r^2$, $r\geqslant R$. Integrating the first equation of (\ref{nadai_3}) and taking the boundary condition $\sigma(R) = -p = const.$ we have
\ben
\sigma = \pm k \ln \left( \frac{r^2}{R^2} + \sqrt{\frac{r^4}{R^4} - 1} \right) - p.
\een
For function $\theta$
\bea
\label{theta_nadai}
\theta = \varphi + \frac{1}{2} \arccos \left(-\frac{R^2}{r^2}\right) = \varphi + \frac{\pi}{2} - \frac{1}{2}\arccos\frac{R^2}{r^2}.
\eea
Finally, taking $\tan g = \left( r^2 / R^2 + 1 \right)/ \sqrt{r^4/R^4 - 1}$ and applying reflection (\ref{reflection3}) to (\ref{theta_nadai}) we obtain the well known vortex flow Nadai solution 
\be
\label{Nadai2}
\sigma = -k\ln{\tan{\left(\frac{1}{2}\arccos\frac{R^2}{r^2} + \frac{\pi}{4} \right)}} - p, \\
\theta = \varphi - \frac{\pi}{2} + \frac{1}{2}\arccos\frac{R^2}{r^2}. 
\ee

The boundary condition for the above solution is $\tau_{r\varphi} |_{r=R} = k\cos 2\theta^p = k \cos2g(R) = -k$, $\sigma|_{r=R} = -p$. The homotopy of the above solution with Prandtl solution ({\ref{Prandtl}}) was analyzed in \cite{Yakhno:2010}.  The generalization of Nadai solution was obtained by Mikhlin \cite{Mikhlin:1938} for $\tau_{r\varphi} = q$, $|q| < k$, when $r=R$. For the corresponding velocity solution see \cite{Sokolovsky:1950}.

b) In the case $\alpha \ne 0$,  the invariants of the operator $\left<\partial_\varphi + \partial_{\theta} + \alpha r\partial_r \right>$ are the following ones: 
\ben
I_1 = z = r e^{-\alpha \varphi},\ I_2 = \sigma,\ I_3 = -\theta + \frac{\varphi}{2} + \frac{1}{2\alpha} \ln r
\een
and the form of the invariant solution is
\ben
\sigma = f(z), \ \theta^c = \frac{\varphi}{2}  + \frac{1}{2\alpha} \ln r = \varphi + \frac{1}{2\alpha} \ln z \sim \theta^p = \frac{1}{2\alpha} \ln z.
\een 
Substitution of this form into system (\ref{polar2}) gives a compatibility condition $\alpha^2 = -1$. That means $z = x \pm i y$ and there is no real solution.

\subsection{$\Theta_4^{(\pm)}$: $\sigma$-translation, rotation and scale}

a) Let $\alpha = 0$, $\Theta_4^{(-)} = \left<\partial_\sigma  - \partial_\varphi - \partial_\theta \right>$. The form of the invariant solution is
\ben
\sigma = - A\varphi + f(r),\ \theta^c = \varphi + g(r), \ A > 0,
\een
which coincides with (\ref{Nadai_two_circ}), taking into account that $g(r) = \theta^p(r)$. The Nadai solution for two concentric circles is an invariant one with respect to operator of $\Theta_4^{(-)}$.

b)  If $\alpha = 0$ and $\Theta_4^{(+)} = \left<\partial_\sigma + \partial_\varphi + \partial_\theta\right>$ is taken, the form of invariant solution is
\ben
\sigma = A\varphi + f(r),\ \theta^c = \varphi + g(r), \ A > 0,
\een
which is dual to solution (\ref{Nadai_two_circ}), (\ref{ff}), (\ref{cc}) in the sense, that families of characteristics change between themselves, and there is no new solution.

c) $\alpha \ne 0$. In such a case, the invariant, relating independent variables is $\lambda = r e^{-\alpha \varphi}$. Taking into account, that $A \partial_\sigma$ is a symmetry,  let us consider the following form of the invariant solution
\be
\label{form_spiral}
\sigma = A \varphi + f(\lambda), \ \theta^c = \varphi + g(\lambda), 
\ee
where $A = const. \ne 0$ and $g = \theta^p$. Let us note, that if $A = \pm 2k$ there is a simple-wave solution, considered later (see paragraph \ref{par:6}).
 
Substituting (\ref{form_spiral}) into (\ref{polar2}) gives two conditions on functions $f$ and $g$:
\ben
f' = \frac{A}{\lambda} \frac{2k/A + \cos 2g - \alpha \sin 2g}{2\alpha \cos 2g + (1-\alpha^2) \sin 2g},  \\
g' = \frac{1}{\lambda} \frac{A/(2k) + \cos 2g - \alpha \sin 2g}{2\alpha \cos 2g + (1-\alpha^2) \sin 2g}, 
\een
which are two separable equations with a closed form solution.

For simplicity let us take $\alpha = 1$. Then function $g$ is defined implicitly by relation 
\ben
\ln \lambda = \int \frac{2\cos 2g}{A/(2k) + \cos 2g - \sin 2g } dg = I_2(g) + const.
\een

Relations (\ref{char_polar}) along characteristic lines take the form:
\ben
\frac{d \ln r}{d \varphi} =  -\tan g, \ \frac{d \ln r}{d \varphi} = \cot g.
\een
But  $\dfrac{d \ln r}{d \varphi} = \dfrac{d \ln \lambda}{d \varphi}  + 1$, so for the first family we have
\ben
r = e^{\varphi + I_2(g) + C_1}, \varphi = I_1(g) + C_1,
\een
and the second family is defined by
\ben
r = e^{\varphi + I_2(g) + C_2}, \varphi = \tilde{I_1}(g) + C_2,
\een
where $C_i$ are the constants of characteristic lines and functions  $I_1$ and $\tilde{I_1}$ are the following ones:
\ben
I_1(g) = -\int \frac{2\cos 2g}{A/(2k) + \cos 2g - \sin 2g } \frac{dg}{\tan g + 1},  \\
\tilde{I_1}(g) = \int \frac{2\cos 2g}{A/(2k) + \cos 2g - \sin 2g } \frac{dg}{\cot g - 1}.
\een

\begin{figure}[h]
\begin{center}
\includegraphics[width=0.5\linewidth]{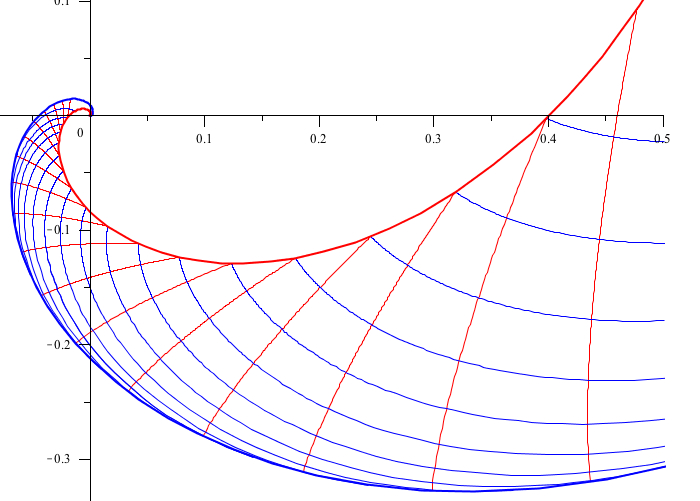}
\caption{Spiral solution}
\label{fig:spiral}
\end{center}
\end{figure}

In Fig. {\ref{fig:spiral}} one can observe two families of  slip-lines and their corresponding envelopes:
\ben
r = \exp \left( \varphi + \frac{\pi - \sqrt{2}}{4} - \frac{1}{2(2-\sqrt{2})} +\frac{1}{4} \ln \frac{\sqrt{2} + 1}{\sqrt{2} - 1} \right),  \\
r = \exp\left(\varphi - \frac{\pi}{8} - \frac{3}{4} \ln 2 \right),
\een
which are two logarithmic spirals. To simplify the above integrals we take $A = 2\sqrt{2} k$, then functions $I_1$, $\tilde{I_1}$ and $I_2$ can be expressed in terms of simple functions. 
 
It seems that the spiral solution was first mentioned by Hartmann \cite{Hartmann:1925}. Unfortunately, this work was not published, but one can find some results in \cite{Nadai:1963}. In works \cite{Annin:1978}, \cite{Annin:1985} the stresses were described. Here we give the corresponding slip-lines field.

\subsection{$\Theta_5$: scale transformation}
\label{par:6}

 The form of invariant solution is 
\ben
\sigma = \sigma(\varphi),\ \theta = \theta(\varphi).
\een
From the first equation of (\ref{polar2}) one can obtain
\ben
\theta - \varphi = \pi n/2,\ n \in \mathbb{Z}.
\een
The second equation gives 
\be
\label{simple_wave}
\sigma = (-1)^n 2k\varphi + const. = (-1)^n 2k \theta^c + const.
\ee

In other words, there are simple wave solutions, which are well known \cite{Kachanov:2004} and which are called simple stress states. Let us note, that in Cartesian coordinates, these solutions correspond to the so-called similarity ones (or dimensionless), i.e. depend on $y/x$ variable only. Unfortunately, it was not seen in \cite{Lamothe:2012}. Solution, obtained by the symmetry reduction for the representative sub-algebra of new symmetry $B_1 = -v\partial_x + u \partial_y$ has the form (\ref{simple_wave}). 

For example, in case $\sigma = 2k \theta + const.$, from (\ref{cartesian}) function $\theta(x,y)$ is defined by the following relation:
\be
\label{simple}
x \cos\theta + y \sin\theta = \Phi(\theta),
\ee 
where $\Phi(\theta)$ is an arbitrary function. Of course, one can take different forms of $\Phi$ (as in \cite{Lamothe:JMATH}, \cite{Lamothe:2012}) to define some solution, but all these solutions correspond to one characteristic family, which is a family of straight-lines $\theta = C_1$:
\be
\label{line}
x \cos C_1 + y \sin C_1 = \Phi(C_1),
\ee
with relation (\ref{char_c}) along characteristic $dy/dx = - \cot C_1$. From (\ref{simple}), taking into account the second relations along characteristic $dy/dx = \tan \theta$, one can obtain linear equation
\ben
\frac{dx}{d\theta} - \cot\theta \left( x + f'(\theta) \sin^2\theta \right) = 0,
\een
where $f(\theta) = \Phi(\theta)/\sin\theta$, with solution
\be
\label{simple_x}
x = \sin\theta \left(\int f' \cos\theta d\theta + \tau \right), \ \tau = const.
\ee
The family of characteristics (\ref{simple}), (\ref{simple_x}) is orthogonal to the family (\ref{line}). Equation 
\be
\label{envelope_simple}
x + f'(\theta) \sin^2\theta = 0
\ee
defines the envelope of straight-lines. All above relations are due to \cite{Khristianovich:1936}.

The velocity fields for simple stress states have some trivial properties \cite{Sokolovsky:1950}. To construct these fields let us use components $U$ and $V$ of the velocity along characteristic directions (\ref{char_cc}):
\ben
u = U \cos\theta - V \sin \theta,\ v = U \sin \theta + V \cos \theta.
\een
The velocity component along each straight line is constant. Thus, in the case of solution (\ref{simple}) $U=const.$ along characteristic and for $V$ from (\ref{uv_cartesian}) we obtain equation
\ben
\frac{\partial V}{\partial x} \sin \theta - \frac{\partial V}{\partial y} \cos \theta = 0,
\een
with general solution $V(x,y) = V(J_1)$, where $V$ is an arbitrary function of $J_1 = x\cos\theta + y\sin \theta$. Finally,
\ben
u = U(\theta) \cos\theta - V(J_1) \sin \theta,\ v = U(\theta) \sin \theta + V(J_1) \cos \theta.
\een

If the second family is a family of straight lines, then by the analogy, we have
\ben
u = U(J_2) \cos\theta - V(\theta) \sin \theta,\ v = U(J_2) \sin \theta + V(\theta) \cos \theta,
\een
where $J_2 = x\sin\theta - y\cos\theta$, and $U$, $V$ are arbitrary functions of their arguments.

\begin{figure}[h]
\centering
\includegraphics[width = 0.5\linewidth]{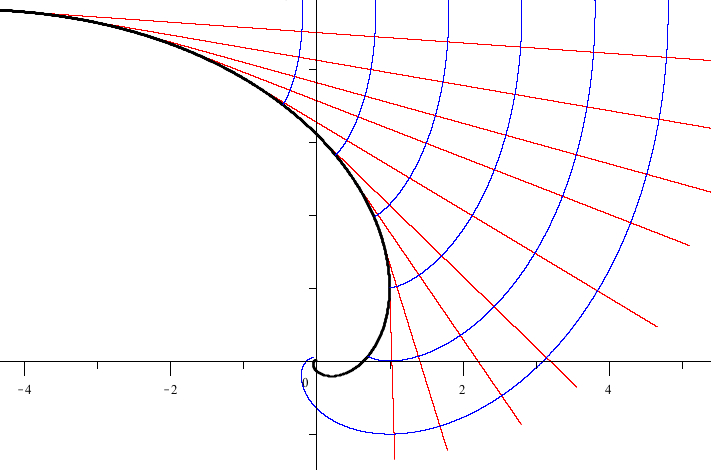}
\caption{Simple wave solution}
\label{fig:spiral_simple}
\end{figure}

Let us note, that the form of the solution, corresponding to simple wave can be quite complicated. For example, one can verify that 
\be
\label{spiral_2}
\sigma = 2k\theta^c + const, \ r\cos(\theta^c - \varphi) = C e^{\theta^c}
\ee
is a solution of (\ref{polar2}). Then, $\Phi(\theta) = C \exp(\theta^c)$ and from (\ref{envelope_simple}) one can obtain a logarithmic spiral
\ben
r = C\sqrt{2} e^{\varphi - \pi/4}
\een
as the envelope for the family $r\cos(C_1 - \varphi) = C e^{C_1}$. The second family of characteristics is defined by the following relations:
\ben
r\cos\varphi = C e^{\theta^c}(\sin\theta^c + \cos\theta^c) + \tau\sin\theta^c, \\
r\sin\varphi = C e^{\theta^c}(\sin\theta^c - \cos\theta^c) - \tau\cos\theta^c.
\een
In Fig. \ref{fig:spiral_simple} one can observe the corresponding slip-line field.

\subsection{Infinite part of symmetries}\label{sub:infinite} Classic Prandtl solution \cite{Prandtl:1923} for system (\ref{equil_c}), (\ref{yield_c})  has the form
\ben
\frac{\sigma_x}{k} = - c - \frac{mx}{h} + 2\sqrt{1-\frac{m^2 y^2}{h^2}}, \\
\frac{\sigma_y}{k} = - c -  \frac{mx}{h},\\
\frac{\tau_{xy}}{k} = \frac{my}{h},
\een
where $h>0$, $0 \leqslant m \leqslant 1$ and $c$ are constants. This solution describes stress state of a thin plastic block of the height $2h$, compressed between rough parallel plates and it satisfies the following boundary conditions
\ben
\tau_{xy}|_{y = \pm h} = \pm mk.
\een

Let us take $c=0$, $2k = 1$, $h=1$ and  $m=1$ (corresponding to perfectly rough plates). Then in terms of functions $\sigma$, $\theta$ we have:
\bea
\label{Prandtl}
 2\sigma = -x + \sqrt{1-y^2}, \ y = \cos 2\theta.
\eea
This solution is invariant with respect to sub algebra of the form $\left<\partial_\sigma - 2 \partial_x\right>$, where  operator $\partial_x$ is of $X_5$ form.

Equations for slip-line families ($\theta$ is the parameter) look as follows:
\ben
 x = \mp 2\theta + \sqrt{1-y^2} + const., \ y = \cos 2\theta,
\een
and represent two orthogonal families of cycloids.

Let us consider the solution, proposed in \cite{Lamothe:JMATH} in a form of propagation wave
\ben
\sigma = f(\xi), \theta = g(\xi), \xi = a_1 x + a_2 y,
\een
where $a_i$ are constants. It is easy to see, that this solution is invariant with respect to the operator $X = a_2\partial_x - a_1\partial_y$, which is of $X_5$ form.  Lie algebra $L_5$ with the basis $\left< X_1, X_2, X_3, \partial_x, \partial_y \right>$, considered in \cite{Annin:1985}, has the non-similar sub-algebra $\Theta_0 = \left< \partial_x + \gamma X_3\right>$, $\gamma \in \mathbb{R}$. If $a_1 \ne 0$, then by the rotation transformation, $a_1 \to 0$ and $X \sim \partial_x \in \Theta_0|_{\gamma = 0}$. That is why one can take $a_1 = 0$, $a_2 = 1$, and the form of invariant solution is $\sigma = f(y)$, $\theta = g(y)$, that gives the trivial constant solution. 

Another invariant, considered  in \cite{Lamothe:JMATH} is $\tau = \sigma + a_1 x + a_2 y$, which corresponds to operator $X = 2 a_1 \partial_\sigma -  \partial_x - a_1/ a_2\, \partial_y$. In the same way we obtain $X\sim 2\partial_\sigma - \partial_x \in \Theta_0|_{\gamma = -2}$, that produces the Prandtl solution (\ref{Prandtl}), so there is no new invariant solution.

\section{Velocity field}\label{sec:velocity}

Lie algebra of operators, admitted by  (\ref{cartesian}), (\ref{uv_cartesian}) is known (see \cite{Annin:1985}, \cite{Lamothe:2012}). Substituting solution $\sigma=\sigma_0(x,y)$, $\theta = \theta_0(x,y)$, one can find the symmetries of (\ref{uv_cartesian}) only. Analyzing the structure of the optimal system of non-similar subalgebras, one can construct corresponding invariant solutions. As an example, let us consider the construction of velocity field for Prandtl solution. 

From (\ref{Prandtl}) we have
\ben
x = -2\sigma - \sin 2\theta, \ y = \cos 2\theta. 
\een

Taking into account that $\partial_\sigma = -2\partial_x$, $\partial_\theta = -2y\partial_x + 2\sqrt{1-y^2}\partial_y$, we have the following Lie algebra $L_{uv}$ of point symmetries:
\ben
Z_1 = u \partial_u + v\partial_v,  \
Z_2 = -2 y \partial_x + 2\sqrt{1-y^2} \,\partial_y - v\partial_u + u \partial_v, \
Z_3 = \partial_x,\nonumber  \\
Z_4 = \left[ 2 \arccos y + 2y\left(x - \sqrt{1-y^2}\right)\right] \partial_x  + 2\left(-x+\sqrt{1-y^2}\right)\sqrt{1-y^2} \partial_y - (xu + yv) \partial_v + \nonumber\\
 + \left[ yu + v\left(x - 2\sqrt{1-y^2}\right) \right]\partial_u,\
Z_5 = y\partial_u - x\partial_v, \nonumber\\
Z_6 = u_0(x,y) \partial_u + v_0(x,y) \partial_v,\nonumber
\een
where $u_0$, $v_0$ is a solution of (\ref{uv_cartesian}). As in the case of Lie algebra $L_{\sigma\theta}$, algebra $L_{uv}$ contains the infnite-dimensional operator $Z_6$, which corresponds to the principle of linear superposition of solutions of linear equations.

Let us note, that $Z_5\sim Z_6$, because $u=y$, $v=-x$ is the trivial solution for any form of $\sigma_0$, $\theta_0$. Symmetry $Z_2$ is an analog of rotation group and $Z_4$ is an analog of quasi-scale transformation.

The non-zero commutators for the finite part of Lie algebra $L_{uv}$ are the following ones:
\be
\label{comm_uv}
[Z_2, Z_4] = - 4 Z_3,\ [Z_3,Z_4] = - Z_2,
\ee
and for infinite part we have
\bea
&&[Z_1, Z_5] = - Z_5, \ [Z_3, Z_5] = -\partial_v, \nonumber\\
&&   [Z_2,Z_5] = \left(-x + 2\sqrt{1-y^2}\right) \partial_u + y \partial_v, \label{Nadai_uv} \\
&& [Z_4, Z_5] =  \left(x^2 -3 y^2 - 4x\sqrt{1-y^2} + 2\right)\partial_u + \left( -2\arccos y - 2xy + 2y \sqrt{1-y^2}\right) \partial_v. \label{yakhno_uv}
\eea

\subsection{Infinite subalgebra}
It is interesting to note, that relation (\ref{Nadai_uv}) produces the well known Nadai solution for velocities \cite{Hill:1950}: 
\be
\label{nadai}
u=x - 2\sqrt{1-y^2}, \ v = -y,
\ee
and from (\ref{yakhno_uv}) we have ($C_i = const.$):
\be
\label{yakhno}
u = x^2 -3 y^2 - 4x\sqrt{1-y^2} + C_1,  \\
v = 2y \sqrt{1-y^2} -  2\arccos y - 2xy + C_2.
\ee
 Using scale transformation $\tilde{u} = a u$, $\tilde{v} = a v$, corresponding to operator $Z_1$, one can multiply the above relation by the same factor $a \ne 0$. And with the help of translations $\partial_u, \partial_v \in Z_6$ we can add any constant to the velocity components.

In general, any combination of commutators of the form $\left[Z_{2,4},\left[Z_{2,4}, \dots \left[Z_{2,4},Z_5 \right]\right.\dots\right]$ produces solutions for velocities. Thus, commutator $\left[Z_{2},\left[Z_4,Z_5 \right]\right]$ gives 
\ben
u= 2xy - 2\arccos y - 2 y \sqrt{1-y^2}, \
v = - x^2 - y^2 + 6,
\een
that coincides with Ivlev-Senashov \cite{Annin:1985} solution of the form
\bea
\label{senashov}
u = xy + \arcsin y - y \sqrt{1-y^2} + const_1, \ v = -(x^2 + y^2)/2 + const_2,
\eea
taking into account that $\arcsin y = \pi/2 - \arccos y$. Let us note, that the above solution is invariant with respect to the operator $\left<Z_3 + Z_5\right>$, and Nadai solution  is the invariant one for the operator $\left<Z_3 + \partial_u\right>$.

Let us give a mechanical interpretation. Nadai solution (\ref{Nadai_uv}) satisfies linear boundary conditions for $u$ on the edges $h=\pm 1$ of the plates
\ben
u(x,\pm 1) = x,\ v = \mp 1,
\een
while $v$ is a constant.

If $C_1 = 3$, $C_2 = \pi$, then solution (\ref{yakhno}) satisfies simple symmetric boundary conditions on the plates $y = \pm 1$:
\ben
u(x,\pm 1) = x^2, \ v(x,\pm 1) = \mp (2x - \pi),
\een
in other words, solution (\ref{yakhno}), in the sense of boundary conditions, is a generalization of Nadai solution, because now for the velocity $u$ we have quadratic dependence on $x$, and for $v$ the dependence is linear.

One of the main properties of plastic deformation is indicated in \cite{Kachanov:2004}: the dissipation must be positive in plastic zone, because a plastic deformation is accompanied by irreversible energy consumption. This is a condition of compatibility of stress and velocity fields. The following condition ensures the non-negativity of the plastic dissipation of energy:
\be
\label{diss_non}
\frac{\dfrac{\partial u}{\partial x} - \dfrac{\partial v}{\partial y}}{\sigma_x - \sigma_y} = \frac{\dfrac{\partial v}{\partial x} + \dfrac{\partial u}{\partial y}}{2\tau_{xy}} \geqslant 0.
\ee
If the above inequality is not satisfied, then boundary conditions are wrong formulated from the mechanical point of view. Thus, for Nadai solution condition (\ref{diss_non}) is satisfied for any $x$: $2/\sqrt{1-y^2} > 0$. For solution (\ref{yakhno}) the plastic dissipation is non-negative if 
\be
\label{plastic_zone}
x \geqslant 2\sqrt{1-y^2},
\ee
i.e. when Nadai solution component $u\geqslant 0$.

To analyze the complete solution of (\ref{cartesian}), (\ref{uv_cartesian}) one needs to know both slip-line field and velocity field. To construct streamlines for the known components of velocity vector field, we can solve equation
\be
\label{flow}
\frac{dx}{u} = \frac{dy}{v}.
\ee 
Thus, for Nadai solution the equation of streamlines is
\ben
x = \frac{y \sqrt{1-y^2}+\arcsin y +const}{y},
\een
and for solution (\ref{yakhno}) it looks as follows
\ben
2 x \arccos y  + y x^2 - 2xy \sqrt{1-y^2} - \pi x - y^3 + 3y = const.
\een

\begin{figure}[h]
\centering
\includegraphics[width=1\linewidth]{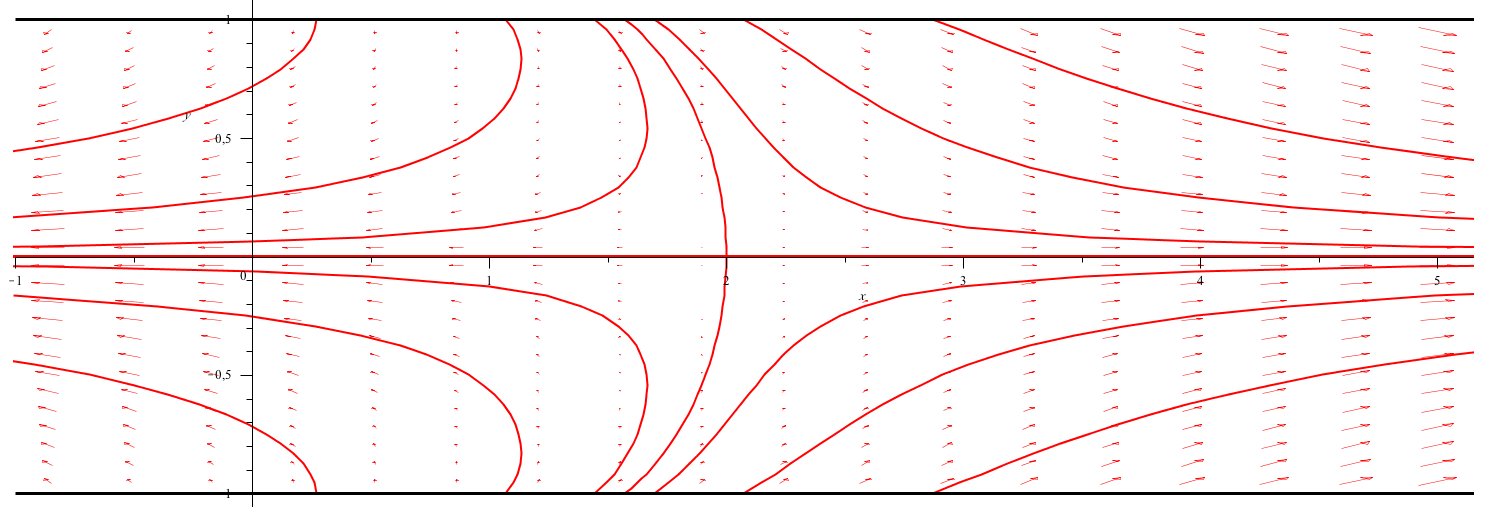}
\caption{Nadai velocity solution for Prandtl stresses}
\label{fig:Nadai_uv}
\end{figure}

\begin{figure}[h]
\centering
\includegraphics[width=1\linewidth]{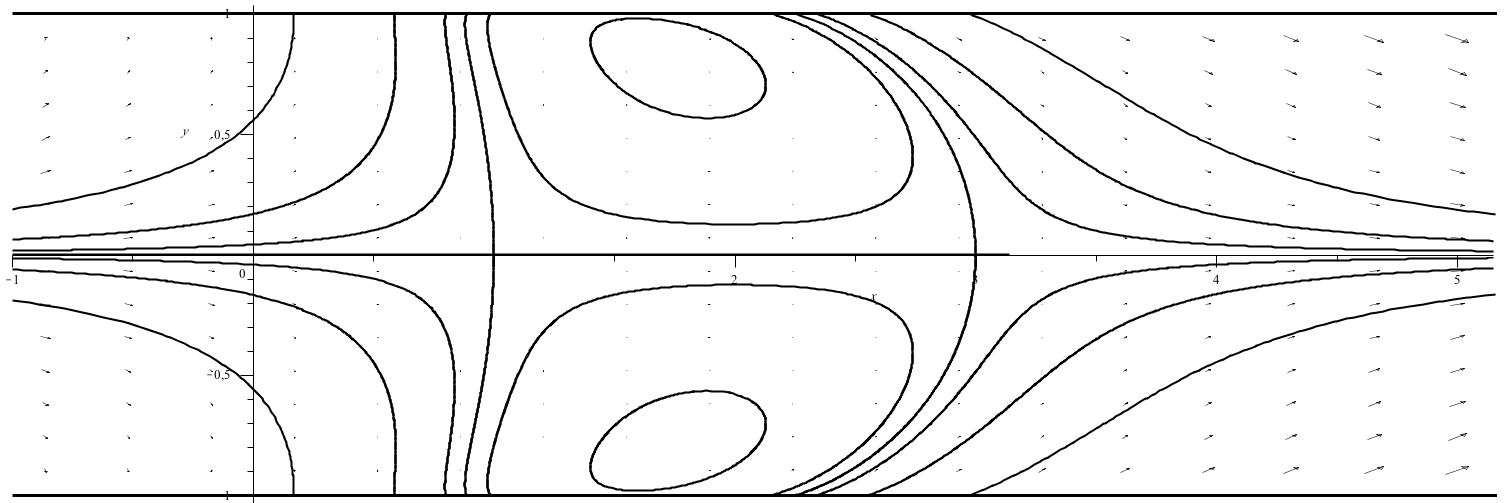}
\caption{Velocity solution (\ref{yakhno}) for Prandtl stresses}
\label{fig:new_uv}
\end{figure}

The theory of plane plastic strain does not involve the time and calculated stresses do not depend on the rate of strain. The progress of the deformation can be expressed by monotonically varying quantity: a load, an angle or, as in case of Prandtl solution, one can take the height of the block (see \cite{Sokolovsky:1950}). Streamlines coincide with the path of material particles and they help to visualize the flow. One can observe vector fields and streamlines for Nadai solution and for solution (\ref{yakhno}) in Figs. \ref{fig:Nadai_uv} and \ref{fig:new_uv} respectively.

Due to linearity of velocity equations there is a infinite number of their solutions. The following condition can be useful for selecting the appropriate one. It is known \cite{Ivlev:1966}, that for the real velocity field, due to modified Mises criterion, the dissipation 
\be
\label{diss}
D = \sigma_x e_x + \sigma_y e_y + 2\tau_{xy} \gamma_{xy},
\ee
where $e_x = \dfrac{\partial u}{\partial x}$, $e_y = \dfrac{\partial v}{\partial y}$, $2\gamma_{xy} = \dfrac{\partial v}{\partial x} + \dfrac{\partial u}{\partial y}$ are corresponding strains, should take a maximum value
\ben
D_{real}\geqslant D_{possible}.
\een

Thus, for Nadai solution $D=1/\sqrt{1-y^2}$ and for solution (\ref{yakhno}) $D = 2 \dfrac{x - 2\sqrt{1-y^2}}{\sqrt{1-y^2}}$, therefore, when $x > 1/2  + 2\sqrt{1-y^2}$, the dissipation of solution (\ref{yakhno}) is greater than dissipation of Nadai solution. 

 Let us note, that Prandtl solution is a good approximation for experimental data only at sufficiently great distance from the centre \cite{Hill:1950}. The same is true for velocity field. 

\subsection{Invariant solutions} Let us consider the finite part of Lie algebra $L_{uv}$. Noting, that commutators (\ref{comm_uv}) coincide with (\ref{comm_sigma}), when $k=1$, one can conclude that the optimal systems are similar:
\ben
\Theta_1 = \left< Z_4 + \alpha Z_1 \right> , \ \Theta_2 = \left< Z_3 + \alpha Z_1 \right> , \Theta_3 = \left< Z_2 + \alpha Z_1 \right>,  \\ 
 \Theta_4^{(\pm)} = \left< Z_3 \pm Z_2 + \alpha Z_1 \right>,  \Theta_5 = \left< Z_1 \right>.
\een

\subsubsection{$\Theta_3$: rotation symmetry}

Let us consider $\alpha = 0$ (only rotation operator) and introduce new unknown functions $\rho(x,y)$, $\psi(x,y)$: 
\be
\label{rho_psi}
u = \rho\sin \psi, \  v = \rho \cos\psi,
\ee
then operator $Z_2$ takes the form
\be
\label{Z2}
\tilde{Z}_2 = -2 y\partial_x + 2\sqrt{1-y^2} \partial_y - \partial_\psi.
\ee
Its invariants are $I_1 = z = x - \sqrt{1-y^2}$, $I_2 =  \psi + \dfrac{1}{2}\arcsin y$, $I_3 = \rho$ and the form of invariant solution is
\ben
\rho = f(z), \ \psi = g(z) - \frac{1}{2} \arcsin y.
\een
Substituting above relations to (\ref{uv_cartesian}) and simplifying, we obtain system
\ben
f g' + \frac{1}{2} f \sin 2g = 0, \ f' - \frac{1}{2} \cos 2g =0.
\een
Taking $g = 1/2\arccos h(z)$ the above system takes the form
\ben
h' = 1- h^2, \ f' = 1/2 f h,
\een
so 
\ben
h = \tanh(z + c_1), \ f = c_2 \cosh^{1/2}(z+c_1), \ \ c_i = const.
\een
and finally
\ben
u = c_2 \cosh^{1/2}(z+c_1) \sin \left( \frac{1}{2} \arccos \tanh(z + c_1) - \frac{1}{2} \arccos y \right), \\
v = c_2 \cosh^{1/2}(z+c_1) \cos \left( \frac{1}{2} \arccos \tanh(z + c_1) - \frac{1}{2} \arccos y \right).
\een

\subsubsection{$\Theta_4$: $x$-translation and rotation}\label{subsec:qscale}

Let us consider the case of $\Theta_4^{(+)}$, when $\alpha = 0$. Then operator of symmetry in terms of variables (\ref{rho_psi}) is slightly different from (\ref{Z2}) and looks as follows
\ben
(1-2y)\partial_x + 2\sqrt{1-y^2} \partial_y - \partial_\psi.
\een
The form of invariant solution is
\ben
\rho = f(z), \ \psi = - \frac{1}{2} \arcsin y + g(z), \ z = x - \sqrt{1-y^2} - \frac{1}{2} \arcsin y.
\een
To determine functions $f$, $g$ by analogy with the above section we come to the system
\ben
\left(\frac{1}{2} - \sin 2 g \right) f' - f g' \cos 2g =0, \
f' \cos 2g - \left(\frac{1}{2} + \sin 2g \right) f g' = f/2,
\een
with solution in quadratures
\ben
g= \arctan\left[2 - \sqrt{3} \tanh \left(\frac{\sqrt{3}}{3} z + const. \right) \right],  \
\ln f = \frac{2}{3} \int \cos 2g(z) dz. 
\een

\subsubsection{$\Theta_2$: $x$ - translation and scale}\label{subsec:scale}

The case $\alpha = 0$ gives a trivial solution $u,v = const.$ Let us consider $\alpha \ne 0$, then invariant solution looks as follows
\be
\label{uv_transl}
u = e^{\alpha x} f(y), \ v = e^{\alpha x} g(y).
\ee
This form was proposed in \cite{Annin:1985}, but the solution was not determined.

For functions $f$, $g$ there is the system:
\ben
\sqrt{1-y^2} \left( f' + \alpha g \right) + y \left( g' - \alpha f \right)  = 0, \ \alpha f + g' = 0.
\een
Eliminating $f$, we obtain the linear equation for $g$
\ben
g^{\prime\prime} - 2\alpha \frac{y}{\sqrt{1-y^2}} g' - \alpha^2 g = 0
\een
with particular solution (when $\alpha = -1/2$) of the form
\ben
g_1(y) = e^{\frac{1}{2} \sqrt{1-y^2}} \left(1 - \sqrt{1-y^2}\right)^{\frac{1}{2}}.
\een
The second particular solution can be found in a well-known way and is as follows
\ben
g_2(y) = \frac{g_1(y)}{y} \left(1 + \sqrt{1-y^2} + y\arcsin y\right).
\een
Finally, solution in the form (\ref{uv_transl}) is
\ben
u(x,y)  = y\, e^{\frac{1}{2}\left(-x+ \sqrt{1-y^2}\right)}  \left(1-\sqrt{1-y^2}\right)^{-\frac{1}{2}}  \left[c_1 + \frac{c_2}{y} \left(-1+\sqrt{1-y^2} + y \arcsin y\right) \right], \\
v(x,y)  =  e^{\frac{1}{2}\left(-x+ \sqrt{1-y^2}\right)}  \left(1-\sqrt{1-y^2}\right)^{\frac{1}{2}}   \left[c_1 +  \frac{c_2}{y} \left(1+\sqrt{1-y^2} + y \arcsin y\right) \right].
\een

In the case when $c_1 \ne 0$, $c_2 = 0$, from equation (\ref{flow}) we obtain a family of streamlines
\ben
x + const. = \sqrt{1-y^2} + \frac{1}{2} \ln\left(\sqrt{1-y^2} - 1\right)^2
\een
and boundary conditions $u(x,\pm 1)=\pm e^{-x/2}$, $v(x,\pm 1) = e^{-x/2}$.

But the case $c_1 = 0$, $c_2 < 0$ seems to be more interesting from mechanical point of view, because in such a case the boundary conditions are: $u(x,\pm 1)=c_2(\pi/2 - 1)e^{-x/2}$, $v(x,\pm 1) = \pm  c_2(1+\pi/2) e^{-x/2}$, i.e. the plates are coming close to one another with the same velocity. 

Moreover, dissipation function (\ref{diss}) has the form
\be
\label{diss_exp}
D = - c_2 \frac{\sqrt{1-y^2} -1 + y \arcsin y}{2 \left(1-\sqrt{1-y^2}\right)^{\frac{1}{2}} \sqrt{1-y^2}}\, e^{\frac{1}{2}\left(-x+ \sqrt{1-y^2}\right)}
\ee
and is non-negative for any $y\in[-1,1]$, so $D \geqslant 0$ along the whole plastic block. 

\begin{figure}[h]
\centering
\includegraphics[width=1\linewidth]{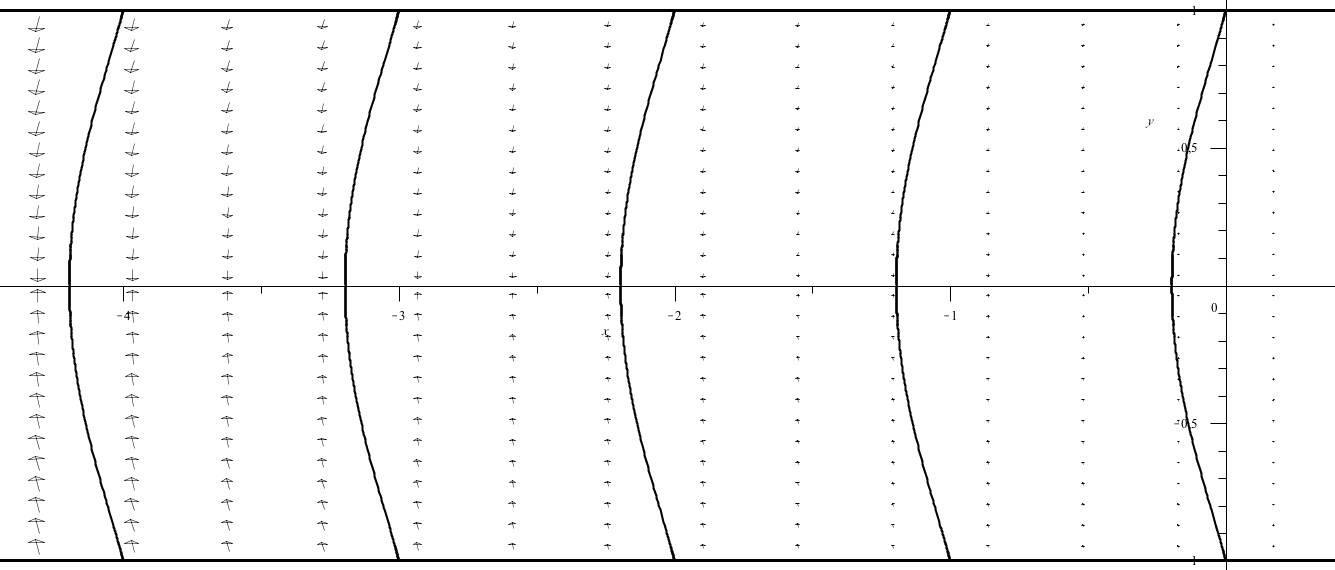}
\caption{Scale velocity solution for Prandtl stresses}
\label{fig:scale_uv}
\end{figure}

The velocity field and a family of streamlines are shown in Fig. (\ref{fig:scale_uv}). Let us note, that streamlines look like in the experimental data, indicated in \cite{Nadai:1950}.

\subsubsection{$\Theta_1$: quasi-scale symmetry}\label{subsec:qqscale}

In the case, when $\alpha = 0$, operator $Z_4$ in terms of variables $\rho$, $\psi$ (\ref{rho_psi}) and $\theta = -1/2 \arccos y$ takes the form
\ben
\tilde{Z}_4 =
\left[- 4\theta + 2 \cos 2\theta (x + \sin 2\theta) \right] \partial_x - (x + \sin 2\theta) \partial_\theta - \cos(2\theta + 2\psi) \rho\partial_\rho + \\
 + \left[ \sin (2\theta + 2\psi) + x + \sin 2\theta \right] \partial_\psi.
\een
The invariant  for independent variables is $z = (x + \sin 2\theta)^2 - 4\theta^2$. But there are some difficulties to find invariants for dependent variables $\rho$, $\theta$, so this case is a matter of the future investigation.

Subalgebra $\Theta_5$ do not produce any form of invariant solution.

\section{Conclusions}

 For the first time, Revuzhenko solution is interpreted as invariant one with respect to quasi-scale transformation. The direct construction of mechanically significant solution, invariant with respect to quasi-scale transformation is quite difficult, because the general solution of the corresponding factor-system of ordinary differential equations is expressed in terms of Bessel functions.  Moreover, the group of symmetry is pointed out, which gives the Nadai solution for two concentric circles.

All of the known classical solutions of plane perfect plasticity system are invariant with respect to some group of point symmetries. One can observe that for different values of the parameters, involved in one non-similar sub-algebra, there are different solutions from mechanical point of view. The equations of slip-line families for all solutions are constructed, which permits to explicitly determine boundaries of plastic areas.

It is shown, how for known stresses one can determine the compatible velocity solution, considering symmetries. It seems there are no advantages in looking for invariants solution using symmetries of complete system (stresses and velocities) in comparison to traditional way exposed in this paper: firstly to solve the system for stresses, substitute its known solution into the system for velocities and determine the invariant velocity solutions. 

As one can see, the streamlines of velocity field even for the simplest Prandtl solution can be different. But the slip-line field is unique and it defines the region of the plastic state, bounded by envelopes of slip-line families (or sometimes by characteristic line). That is why, when one gives a mechanical interpretation of the obtained solution it is necessary to consider both fields simultaneously.   

For the Prandtl solution the well known Nadai velocity solution is compared to some other solutions. New velocity field in the form of exponential function is determined. It is shown, that Nadai velocity solution is not preferable in the whole plastic area. There are other velocities fields, where the dissipation is greater. Due to infinite number of solutions, the problem of the construction of real velocities for Prandtl solutions is still open. 

\section*{Acknowledgements}

The work was partially supported by PRO-SNI 2013 (UDG).

\end{document}